\documentclass[12pt]{article}
\usepackage{amssymb}
\usepackage{epsfig}

\catcode`\@=11
\def\numberbysection{\@addtoreset{equation}{section}
        \def\theequation{\thesection.\arabic{equation}}}
\def\half{\frac{1}{2}}
\def\beq{\begin{equation}}
\def\eeq{\end{equation}}
\numberbysection
\begin{document}
\begin{titlepage}
\begin{center}
\hfill DFF  1/9/02 \\
\vskip 1.in {\Large \bf Solitons for the fuzzy sphere from matrix
model} \vskip 0.5in P. Valtancoli
\\[.2in]
{\em Dipartimento di Fisica, Polo Scientifico Universit\'a di Firenze \\
and INFN, Sezione di Firenze (Italy)\\
Via G. Sansone 1, 50019 Sesto Fiorentino, Italy}
\end{center}
\vskip .5in
\begin{abstract}
Recently we have introduced a matrix model depending on two
coupling constants $g^2$ and $\lambda$, which contains the fuzzy
sphere as a background; to obtain the classical limit $g^2$ must
depend on $N$ in a precise way. In this paper we show how to
obtain the classical solitons of the $N \rightarrow \infty$ limit
imposing the development $\lambda = \half + \frac{\lambda_0}{N}$;
as a consequence at finite $N$ one obtains a noncommutative
version of the solitons for the fuzzy sphere.
\end{abstract}
\medskip
\end{titlepage}
\pagenumbering{arabic}
\section{Introduction}

Recently, noncommutative gauge theories \cite{1}-\cite{2}-\cite{3}
on a noncommutative sphere have been studied by expanding a matrix
model \cite{8}-\cite{9}-\cite{10}-\cite{11}-\cite{12}-\cite{13}
around its classical solution
\cite{4}-\cite{5}-\cite{6}-\cite{7}-\cite{32}. The fuzzy sphere is
considered as a classical background, and the fluctuations on the
background are the fields of noncommutative gauge theory. In a
previous paper \cite{32} we have introduced a matrix model which
depends on two coupling constants, $g^2$ and $\lambda$:

\beq S(\lambda) = S_0 + \lambda S_1 = - \frac{1}{g^2} Tr [
\frac{1}{4} [A_i,A_j][A^i,A^j] - \frac{2}{3}i\lambda \rho
\epsilon^{ijk} A_i A_j A_k + \rho^2 ( 1-\lambda ) A^i A_i ].
\label{11} \eeq

We looked for other classical solutions besides the fuzzy sphere
and found them for $\lambda < 1$. These make the fuzzy sphere
solution unstable for $\lambda < \half$ and stable otherwise.

In this paper we answer a question left in the previous one, i.e.
whether this new class of classical solutions leads to solitons
for the noncommutative gauge theory on the fuzzy sphere.

Generally a solution of the matrix model could be defined as a
soliton, but we have to be more restrictive, i.e. to call a
noncommutative soliton a configuration which tends to a classical
soliton of the sphere in the $N \rightarrow \infty$ limit.

What is called fluctuation is the difference between the fuzzy
sphere $\hat{x}^i$ and the new class of classical solutions $A^i$,
but to recover a well-defined classical limit, the fluctuations
must go to zero as $\frac{1}{N}$, in the $N \rightarrow \infty$
limit \cite{4}-\cite{6}. Such property distinguishes what can be
called a field of noncommutative gauge theory from something
indefinite. Therefore many of the classical solutions of the
matrix model are meaningless, also if they can make the fuzzy
sphere background unstable \cite{32}.

To find a good fluctuation, we need to explore the neighborhood of
$\lambda = \half$ ( precisely $\lambda = \half +
\frac{\lambda_0}{N}$ ), since then our new class of classical
solutions is a continuous deformation of the fuzzy sphere, and the
difference between them is of order $\frac{1}{N}$.

It was already clear from \cite{4}-\cite{6} that the coupling
constant $g^2$ must scale in a precise way with $N$ to recover a
classical gauge theory on a sphere; the idea of the present
article is to introduce a double scaling limit, for both $g^2$ and
$\lambda$, to recover the classical solitons on the sphere from
the classical solutions of the matrix model.

To verify that we have found the noncommutative solitons of the
fuzzy sphere, we check that the gauge field and the scalar field
defined from the fluctuations, obtained from the difference of two
solutions of the matrix model, indeed satisfy the classical
equations of motion, in the $N \rightarrow \infty$ limit.

What we find is that, as typical for the case of solitons, an
equation of motion for the gauge field is already present into the
action as a quadratic term, while the other equation for the
scalar field is equal to the eigenvalue problem for the $L^2$
operator. The only possible solutions of this equation are the
spherical harmonics $Y_l^m (\Omega)$, and in our case it turns out
that the scalar field is a combination of $Y_l^m ( \Omega )$ with
$l=2$. To find noncommutative solitons with higher spherical
harmonics $l>2$, we conclude that one should explore models with $
\lambda = \frac{3-l}{2} $. By analogy with the $\lambda = \half$
case treated here, one should start from solutions of matrix model
obtained with more general ansatze

\beq A_i + A_i^{j_1,...,j_{l-1}} L_{j_1}...L_{j_{l-1}} \label{12}
\eeq

hoping that, around $ \lambda = \frac{3-l}{2} $ , these solutions
are smooth deformations of the fuzzy sphere. To conclude our
analysis, we  evaluate  the action of the matrix model at $\lambda
= \half + \frac{\lambda_0}{N}$, and compare it with the value of
the action of the gauge theory on the sphere, at $\lambda =
\half$, and find complete agreement.

\section{Solutions of the matrix model for $\lambda = \half +
\frac{\lambda_0}{N}$ }

In a recent paper \cite{32} we have introduced an action depending
on two coupling constants, $g^2$ and $\lambda$, which contains as
a solution the fuzzy sphere

\beq S(\lambda) = S_0 + \lambda S_1 = - \frac{1}{g^2} Tr [
\frac{1}{4} [A_i,A_j][A_i,A_j] - \frac{2}{3} i \rho \epsilon^{ijk}
A_i A_j A_k + \rho^2 ( 1-\lambda ) A_i A_i ]. \label{21} \eeq

The fuzzy sphere
\cite{14}-\cite{15}-\cite{16}-\cite{17}-\cite{18}-\cite{19} is a
noncommutative manifold represented by the following algebra:

\beq [ \hat{x}_i, \hat{x}_j ] = i \rho \epsilon^{ijk} \hat{x}_k \
\ \ \ \ \ \ \ \ \hat{x}^i = \rho L^i. \label{22} \eeq

The radius of the sphere, obtained by the following condition,

\beq \hat{x}_i \hat{x}_i = R^2 = \rho^2 L^i L^i = \rho^2
\frac{N(N+2)}{4} \label{23} \eeq

is kept fixed in the commutative limit $N\rightarrow \infty$,
therefore

\beq \rho \sim \frac{1}{N}. \label{24} \eeq

The equations of motion for the matrix model action $S(\lambda)$
contain other solutions rather than the fuzzy sphere, as found in
\cite{32}. Our aim is to use these solutions to construct solitons
solutions for the fuzzy sphere
\cite{20}-\cite{21}-\cite{22}-\cite{23}-\cite{31}. Let us recall
them, starting from the equations of motion

\beq [ A^j, [A^i, A^j]] - i \rho \lambda \epsilon^{ijk} [ A^j, A^k
] + 2 \rho^2 ( 1 - \lambda ) A^i = 0. \label{25} \eeq

Parameterizing the solution as

\beq A^i = A^i_k L^k \label{26} \eeq

the coefficients $A^i_k$ represent three arbitrary vectors $A^1_i,
A^2_i, A^3_i$ that are constrained by the equations

\begin{eqnarray}
& \ & ( 2 \rho^2 ( 1- \lambda ) - \beta^2 - \gamma^2 ) A^1_i + (
A^1 \cdot A^2 ) A^2_i + ( A^1 \cdot A^3 ) A^3_i + 2 \lambda \rho
\epsilon_{ijk} A^2_j A^3_k = 0 \nonumber \\
& \ & ( 2 \rho^2 ( 1- \lambda ) - \alpha^2 - \gamma^2 ) A^2_i + (
A^1 \cdot A^2 ) A^1_i + ( A^2 \cdot A^3 ) A^3_i + 2 \lambda \rho
\epsilon_{ijk} A^3_j A^1_k = 0 \nonumber \\
& \ & ( 2 \rho^2 ( 1- \lambda ) - \alpha^2 - \beta^2 ) A^3_i + (
A^1 \cdot A^3 ) A^1_i + ( A^2 \cdot A^3 ) A^2_i + 2 \lambda \rho
\epsilon_{ijk} A^1_j A^2_k = 0 \label{27}
\end{eqnarray}

where

\beq (A^1_i)^2 = \alpha^2 \ \ \ \ \ (A^2_i)^2 = \beta^2 \ \ \ \ \
(A^3_i)^2 = \gamma^2. \label{28} \eeq

Without loss of generalities, we can choose the three vectors as
follows:

\begin{eqnarray}
& \ & A^1_x = \alpha \ \ \ \ \ \ A^1_y = A^1_z = 0 \nonumber \\
& \ & A^2_x = \beta cos \theta_{12} \ \ \ \ \ A^2_y = \beta sin
\theta_{12} \ \ \ \ \ A^2_z = 0 \nonumber \\
& \ & A^3_x = \gamma cs \theta_{13} \ \ \ \ \ A^3_y = \gamma sin
\theta_{13} sin \phi \ \ \ \ \ A^3_z = \gamma sin \theta_{13} cos
\phi. \label{29}
\end{eqnarray}

Apart from the fuzzy sphere, there is another class of solutions
to the equations (\ref{27}), characterized by the unique
constraint:

\beq \alpha^2 + \beta^2 + \gamma^2 = 4 \rho^2 ( 1-\lambda ) + 4
\lambda^2 \rho^2 \label{210} \eeq

and by the parameterizations

\begin{eqnarray}
& \ & cos \phi = - \frac{ 2 \lambda \alpha \rho}{\sqrt{ ( 2\rho^2
( 1-\lambda ) - \beta^2 )( 2 \rho^2 ( 1-\lambda ) -  \gamma^2 ) +
4 \lambda^2 \alpha^2 \rho^2 }} \nonumber \\
& \ & sin \phi = \sqrt{ \frac{ ( 2 \rho^2 ( 1- \lambda ) - \beta^2
)( 2 \rho^2 ( 1- \lambda ) - \gamma^2 )}{ ( 2 \rho^2 ( 1- \lambda
) - \beta^2 )( 2 \rho^2 ( 1- \lambda ) - \gamma^2 ) + 4 \lambda^2
\alpha^2 \rho^2 }} \nonumber \\
& \ & cos \theta_{12} = - \frac{1}{\alpha\beta} \sqrt{( 2 \rho^2 (
1- \lambda ) - \alpha^2 )( 2 \rho^2 ( 1- \lambda ) - \beta^2 )}
\nonumber \\
& \ & sin \theta_{12} = \frac{1}{\alpha\beta} \sqrt{ 2 \rho^2 ( 1
- \lambda ) ( \alpha^2 + \beta^2 - 2 \rho^2 ( 1 - \lambda ) ) }
\nonumber \\
& \ & cos \theta_{13} = - \frac{1}{\alpha\gamma} \sqrt{ ( 2 \rho^2
( 1- \lambda ) -\alpha^2 ) ( 2 \rho^2 ( 1- \lambda ) - \gamma^2 )
} \nonumber \\
& \ & sin \theta_{13} = - \frac{1}{\alpha\gamma} \sqrt{ 2 \rho^2 (
1- \lambda ) ( \alpha^2 + \gamma^2 - 2 \rho^2 ( 1-\lambda ) }
\nonumber \\
& \ & cos \theta_{23} = - \frac{1}{\beta\gamma} \sqrt{ 2 \rho^2 (
1- \lambda ) - \beta^2 )( 2 \rho^2 ( 1 - \lambda ) - \gamma^2 ) }
\nonumber \\
& \ & sin \theta_{23} = - \frac{1}{\beta\gamma} \sqrt{ 2 \rho^2 (
1 - \lambda ) ( \beta^2 + \gamma^2 - 2 \rho^2 ( 1- \lambda ) ) }
\label{211}
\end{eqnarray}

where $ cos \theta_{23} = cos \theta_{12} cos \theta_{13} + sin
\theta_{12} sin \theta_{13} sin \phi $.

The basic requirement to construct solitons solutions is
developing the matrix $A^i$ as a background plus fluctuations :

\beq A_i = \hat{x}_i + \rho R \hat{a}_i. \label{212} \eeq

In front of the $U(1)$ noncommutative connection $\hat{a}_i$ there
is a factor $\rho$, which means that in the $N \rightarrow \infty$
limit the difference between $A_i$ and the background $\hat{x}_i$
must be negligible. Instead the general class of solutions
(\ref{29}), (\ref{210}), (\ref{211}) $A_i$ differs significantly
from the fuzzy sphere $\hat{x}_i$ and cannot be used to define a
consistent connection as in (\ref{212}).

However we find fruitful exploring the neighborhood of
$\lambda=\half$, since the difference of the classical action
computed on the class of solutions (\ref{211}) and the classical
action computed on the fuzzy sphere goes like ${(\lambda -
\half)}^3$, and can be negligible in the $N \rightarrow \infty$
limit if $\lambda$ scales as

\beq \lambda = \half + \rho \lambda_0. \label{213} \eeq

It is aim of this paper to show that this further constraint can
be implemented to give non trivial connections $\hat{a}_i$, which
then describe noncommutative solitons for the fuzzy sphere.

The first difficulty we meet is to find compatibility between the
two constraints

\begin{eqnarray}
& \ & \alpha^2 + \beta^2 + \gamma^2 = 4 \rho^2 ( 1-\lambda ) + 4
\rho^2 \lambda^2 \nonumber \\
& \ & \lambda = \half + \rho \lambda_0 \label{214}
\end{eqnarray}

since the general solution for $A^1_i$, $A^2_i$ and $A^3_i$ is
constructed uniquely in terms of $\alpha, \beta , \gamma$.

To find compatibility, we need to parameterize $\alpha, \beta,
\gamma$ as:

\begin{eqnarray}
& \ & \alpha = \rho + \alpha_0 \rho^2 \nonumber \\
& \ & \beta = \rho + \beta_0 \rho^2 \nonumber \\
& \ & \gamma = \rho + \gamma_0 \rho^2 \label{215}
\end{eqnarray}

from which it follows that

\begin{eqnarray}
& \ & \alpha_0 + \beta_0 + \gamma_0 = 0 \nonumber \\
& \ & \alpha^2_0 + \beta^2_0 + \gamma^2_0 = 4 \lambda^2_0.
\label{216}
\end{eqnarray}

A general solution of this system is given by:

\begin{eqnarray}
& \ & \alpha_0 = 2 \sqrt{\frac{2}{3}} \lambda_0 sin (
\frac{\pi}{3} -
\theta ) \nonumber \\
& \ & \beta_0 = 2 \sqrt{\frac{2}{3}} \lambda_0 sin \theta \nonumber \\
& \ & \gamma_0 = - 2 \sqrt{\frac{2}{3}} \lambda_0 sin (
\frac{\pi}{3} + \theta ). \label{217}
\end{eqnarray}

This is not the whole story, since the general solution
(\ref{211}), to be well defined, must satisfy the following
inequalities

\beq \alpha^2 \geq 2\rho^2 ( 1- \lambda ) \ \ \ \ \ \beta^2 \geq 2
\rho^2 ( 1-\lambda ) \ \ \ \ \ \gamma^2 \geq 2 \rho^2 ( 1-\lambda
). \label{218} \eeq

By defining

\beq sin \alpha_0 = \half \sqrt{\frac{3}{2}} \label{219} \eeq

these inequalities imply that

\beq sin ( \frac{\pi}{3} - \theta ) \geq - sin \alpha_0 \ \ \ \ \
sin \theta \geq - sin \alpha_0 \ \ \ \ \ sin ( \frac{\pi}{3} +
\theta ) \leq sin \alpha_0. \label{220} \eeq

It is surprising to observe that there is a narrow window of
values for $\theta$ which satisfies all these inequalities:

\beq \frac{4}{3} \pi - \alpha_0 < \theta < \pi + \alpha_0.
\label{221} \eeq

In the following we will use a simple fixed value of $\theta$ to
simplify the calculations :

\beq \theta = \frac{7}{6} \pi \label{222} \eeq

and the general solution to the constraints (\ref{216}) and
(\ref{220}) is determined to be :

\begin{eqnarray}
& \ & \alpha = \rho ( 1 - \sqrt{\frac{2}{3}} \lambda_0 \rho ) \nonumber \\
& \ & \beta = \rho ( 1 - \sqrt{\frac{2}{3}} \lambda_0 \rho )
\nonumber \\
& \ & \gamma = \rho ( 1 + 2 \sqrt{\frac{2}{3}} \lambda_0 \rho ).
\label{223}
\end{eqnarray}

\section{Solitons for the fuzzy sphere }

In order to determine the solitons for the fuzzy sphere we have to
recall how to recover gauge theory on the fuzzy sphere, by
expanding the $A_i$ matrices around the classical solution
(\ref{22}) as:

\beq A_i = \hat{x}_i + \rho R \hat{a}_i. \label{31} \eeq

The action $S(\lambda)$ (\ref{21}) is invariant under the unitary
transformation

\beq A_i \rightarrow U^{-1} A_i U \label{32} \eeq

which implements the gauge symmetry of the noncommutative gauge
theories as a global symmetry of the matrix model.

In fact, by developing $U$ in terms of an infinitesimal
transformation

\beq U \sim 1 + i \hat{\lambda} \label{33} \eeq

the fluctuations around the fixed background transforms as

\beq \hat{a}_i \rightarrow \hat{a}_i - \frac{i}{R} [ \hat{L}_i,
\hat{\lambda} ] + i [ \hat{\lambda}, \hat{a}_i ]. \label{34} \eeq

The corresponding field strength on the sphere is given by

\begin{eqnarray} \hat{F}_{ij} & = & \frac{1}{\rho^2 R^2} ( [ A_i, A_j ] - i \rho
\epsilon_{ijk} A_k ) \nonumber \\
& = & [ \frac{\hat{L}_i}{R}, \hat{a}_j ] - [ \frac{\hat{L}_j}{R},
\hat{a}_i ] + [ \hat{a}_i, \hat{a}_j ] - \frac{i}{R}
\epsilon_{ijk} \hat{a}_k . \label{35} \end{eqnarray}

$ \hat{F}_{ij} $ is gauge covariant even in the $U(1)$ case, as it
is manifest from the viewpoint of the matrix model.

The model contains also a scalar field which belongs to the
adjoint representation as the gauge field and that can be defined
as

\beq \hat{\phi} = \frac{1}{2 \rho R } ( A_i A_i - \hat{x}_i
\hat{x}_i ) = \half ( \hat{x}_i \hat{a}_i + \hat{a}_i \hat{x}_i +
\rho R \hat{a}_i \hat{a}_i ) . \label{36} \eeq

At the noncommutative level, the scalar model is intrinsically
connected with the gauge field and only in the classical limit the
action can be interpreted as a sum of both contributions.

To define the action in terms of the fluctuations $\hat{a}_i$ , we
have to define a star product on the fuzzy sphere analogous to the
Moyal star product for the plane.

Recall that a matrix on the fuzzy sphere can be developed in terms
of the noncommutative analogue of the spherical harmonics
$\hat{Y}_{lm}$:

\beq \hat{Y}_{lm} = R^{-l} \sum_a \ f_{a_1, a_2, ... a_l}^{(lm)}
\hat{x}_{a_1} ... \hat{x}_{a_l} \label{37} \eeq

while the classical spherical harmonics are defined with
$\hat{x}_i$ substituted with the commutative coordinates $x_i$.

A general matrix

\beq \hat{a} = \sum^{N}_{l=0} \sum^{l}_{m=-l} a_{lm} \hat{Y}_{lm}
\ \ \ \ \ a_{lm}^{*} = a_{l -m} \label{38} \eeq

corresponds therefore to an ordinary function on the commutative
sphere as:

\beq a( \Omega ) = \frac{1}{N+1} \sum^{N}_{l=0} \sum^{l}_{m=-l} Tr
( \hat{Y}^{\dagger}_{lm} \hat{a} ) Y_{lm} ( \Omega ) \label{39}
\eeq

and the ordinary product of matrices is mapped to the star product
on the commutative sphere

\begin{eqnarray}
& \ & \hat{a} \hat{b}  \rightarrow a * b \nonumber \\
& \ & a( \Omega ) * b( \Omega ) = \frac{1}{N+1} \sum^{N}_{l=0}
\sum^{l}_{m=-l} Tr ( \hat{Y}^{\dagger}_{lm} \hat{a} \hat{b} )
Y_{lm} ( \Omega ) . \label{310}
\end{eqnarray}

Derivative operators can be constructed using the adjoint action
of $\hat{L}_i$ and tend to the classical Lie derivative $L_i$ in
the $N \rightarrow \infty$ limit:

\beq Ad ( \hat{L}_i ) \rightarrow L_i = \frac{1}{i} \epsilon_{ijk}
x_j \partial_k . \label{311} \eeq

$L_i$ can be expanded in terms of the Killing vectors of the
sphere

\beq L_i = - i K_i^a \partial_a . \label{312} \eeq

In terms of $K_i^a$ we can form the metric tensor $g_{ab} = K^i_a
K^i_b$. The explicit form of these Killing vectors is
\begin{eqnarray}
& \ & K_1^{\theta} = - sin \phi \ \ \ \ \ K_1^{\phi} = - cotg
\theta cos \phi \nonumber \\
& \ & K_2^{\theta} = cos \phi \ \ \ \ \ K_2^{\phi} = - cotg
\theta sin \phi \nonumber \\
& \ & K_3^{\theta} = 0 \ \ \ \ \ K_3^{\phi} = 1  . \label{313}
\end{eqnarray}

Trace over matrices can be mapped to the integration over
functions:

\beq \frac{1}{N+1} Tr ( \hat{a} ) \rightarrow \int
\frac{d\Omega}{4\pi} a ( \Omega ) . \label{314} \eeq

Having introduced the star product, we can compute the action
$S(\lambda)$ as the following field theory action

\begin{eqnarray}
& \ & S(\lambda) = S_0 + \lambda S_1 \nonumber \\
& \ & S_0 = - \frac{ R^2}{4 g^2_{ym}} Tr \int d\Omega ( F_{ij}
F_{ij} ) - \frac{3i}{2 g^2_{ym}} \epsilon_{ijk} Tr \int d\Omega (
(L_i a_j)a_k + \frac{R}{3} [a_i, a_j]a_k - \frac{i}{2}
\epsilon_{ijl} a_l a_k )_{*} \nonumber \\
& \ & - \frac{\pi}{g^2_{ym}} \frac{N(N+2)}{2R^2} \nonumber \\
& \ & S_1 = \frac{i}{2 g^2_{ym}} \epsilon_{ijk} Tr \int d\Omega (
(L_i a_j)a_k + \frac{R}{3} [a_i, a_j]a_k - \frac{i}{2}
\epsilon_{ijl} a_l a_k )_{*} \nonumber \\
& \ & + \frac{\pi}{3 g^2_{ym}} \frac{N(N+2)}{2R^2} \label{315}
\end{eqnarray}

where the residual Trace is in general for the $U(n)$ case,
defined as in \cite{4}-\cite{6}.

Our solution $A_i$, defined by the equations (\ref{223}) gives
rise to a solution of the action $S(\lambda)$ and therefore is a
soliton solution for the fuzzy sphere, since the fluctuations
respect the dependence on $\rho$ of equation (\ref{212}). However
to help intuition we will look for the classical limit of this
solution and verify that it is a nontrivial solution of the
classical limit of the action $S(\lambda)$, i.e. a classical
soliton for the sphere.

The classical limit is realized as

\beq R = {\rm fixed} \ \ \ \ \ g^2_{ym} = \frac{4\pi g^2 }{(N+1)
\rho^4 R^2} = {\rm fixed} \ \ \ \ N \rightarrow \infty .
\label{316} \eeq

In the commutative limit, the star product becomes the commutative
product. In this limit, the scalar field $\phi$ and the gauge
field are separable from each other as in

\beq R a_i ( \Omega) = K_i^a b_a ( \Omega ) + \frac{x_i}{R} \phi(
\Omega ) \label{317} \eeq

where $b_a$ is a gauge field on the sphere. The field strength
$F_{ij}$ can be expanded in terms of the gauge field $b_a$ and the
scalar field $\phi$ as follows

\beq F_{ij} ( \Omega ) = \frac{1}{R^2} K^a_i K^b_j F_{ab} +
\frac{i}{R^2} \epsilon_{ijk} x_k \phi + \frac{1}{R^2} x_j K^a_i
D_a \phi - \frac{1}{R^2} x_i K_j^a D_a \phi \label{318} \eeq

where $F_{ab} = - i ( \partial_a b_b - \partial_b b_a ) + [ b_a,
b_b ] $ and $ D_a = -i \partial_a + [ b_a, ... ] $, in general for
the $U(n)$ case.

Let us identify the classical fluctuations $a_i$ in terms of the
solution for $A_i$ (\ref{223}), developed in power of $\rho$:

\begin{eqnarray}
A^1_x & = & \rho ( 1 - \sqrt{\frac{2}{3}} \lambda_0 \rho )
\nonumber \\
A^2_x & = & - 2 \lambda_0 ( 1 - \sqrt{\frac{2}{3}} ) \rho^2 ( 1 -
\sqrt{\frac{2}{3}} \lambda_0 \rho ) + O( \rho^4 ) \nonumber \\
A^2_y & = & \rho ( 1 - \sqrt{\frac{2}{3}} \lambda_0 \rho - 2
\lambda_0^2 {( 1 - \sqrt{\frac{2}{3}} )}^2 \rho^2 ) + O(\rho^4)
\nonumber \\
A^3_x & = & - 2 \lambda_0 \sqrt{( 1 - \sqrt{\frac{2}{3}} )( 1 + 2
\sqrt{\frac{2}{3}} )} \rho^2 ( 1 + 2 \sqrt{\frac{2}{3}} \lambda_0
\rho ) + O ( \rho^4 ) \nonumber \\
A^3_y & = & 2 \lambda_0 \sqrt{( 1 - \sqrt{\frac{2}{3}} )( 1 + 2
\sqrt{\frac{2}{3}} )} \rho^2 ( 1 + 2 \sqrt{\frac{2}{3}} \lambda_0
\rho ) + O ( \rho^4 ) \nonumber \\
A^3_z & = & \rho ( 1 + 2 \sqrt{\frac{2}{3}} \lambda_0 - 4
\lambda_0^2 ( 1 - \sqrt{\frac{2}{3}} )( 1 + 2 \sqrt{\frac{2}{3}}
)\rho^2 ) + O (\rho^4) . \label{319} \end{eqnarray}

We have kept the first subleading contribution to the classical
fluctuation $a_i$, since , as we will see in the next section,
they can in principle give a finite contribution to the action
$S(\lambda)$. The classical fluctuations $a_i$ are deducible from
(\ref{319}) as follows:

\begin{eqnarray}
R a^1 & = & - \sqrt{\frac{2}{3}} \lambda_0 x^1 \nonumber \\
R a^2 & = & - 2 \lambda_0 ( 1 - \sqrt{\frac{2}{3}} ) x_1 -
\sqrt{\frac{2}{3}} \lambda_0 x_2 \nonumber \\
R a_3 & = & - 2 \lambda_0
\sqrt{(1-\sqrt{\frac{2}{3}})(1+2\sqrt{\frac{2}{3}})} ( x_1 - x_2)
+ 2 \sqrt{\frac{2}{3}}\lambda_0 x_3 . \label{320}
\end{eqnarray}

The classical scalar field, obtained from this particular
fluctuation $a_i$ , is :

\beq R \phi( \Omega ) = x_i \cdot R a_i \label{321} \eeq

therefore

\begin{eqnarray}
R \phi( \Omega ) & = & \sqrt{\frac{2}{3}} \lambda_0 (x^1)^2 - 2
\lambda_0 ( 1 - \sqrt{\frac{2}{3}} ) x^1 x^2 - \sqrt{\frac{2}{3}}
( x^2 )^2 \nonumber \\
& - & 2 \lambda_0 \sqrt{ ( 1- \sqrt{\frac{2}{3}} ) ( 1 + 2
\sqrt{\frac{2}{3}} )} ( x^1 - x^2 ) x^3 \nonumber \\
& + & 2 \sqrt{\frac{2}{3}} \lambda_0 ( x^3 )^2 \label{322}
\end{eqnarray}

or, in usual spherical coordinates,

\begin{eqnarray} \phi( \Omega ) & = & R [ \lambda_0 \sqrt{\frac{2}{3}} ( 2 cos^2
\theta - sin^2 \theta ) - 2 \lambda_0 ( 1 - \sqrt{\frac{2}{3}} )
sin^2 \theta sin \phi cos \phi \nonumber \\
& + & 2 \lambda_0 \sqrt{ ( 1 - \sqrt{\frac{2}{3}} ) ( 1 + 2
\sqrt{\frac{2}{3}} ) } sin \theta cos \theta ( sin \phi - cos \phi
) ] . \label{323} \end{eqnarray}

From the explicit form of the Killing vector $K_i^a$, we can
deduce the two components of the $U(1)$ gauge field as:

\begin{eqnarray}
b_{\theta} & = & - R a^1 sin \phi + R a^2 cos \phi = \nonumber \\
& = & - 2 R \lambda_0 ( 1- \sqrt{\frac{2}{3}} ) sin \theta cos^2
\phi \label{324} \end{eqnarray}

and

\begin{eqnarray}
b_{\phi} & = & R [ 2 \lambda_0 \sqrt{ ( 1- \sqrt{\frac{2}{3}} ) (
1 + 2 \sqrt{\frac{2}{3}} ) } sin^3 \theta ( sin \phi - cos \phi )
\nonumber \\
& + & 3 \lambda_0 \sqrt{\frac{2}{3}} cos \theta sin^2 \theta + 2
\lambda_0 ( 1 - \sqrt{ \frac{2}{3} } ) sin^2 \theta cos \theta sin
\phi cos \phi ] . \label{325}
\end{eqnarray}

Having made contact with the soliton solution, we now verify that
it is a non trivial solution of the classical equations of motion
on the sphere.

The classical limit of the action $S(\lambda)$ is determined to
be:

\begin{eqnarray}
S(\lambda) & =  & S_0 + \lambda S_1 \nonumber \\
S_0 & = & - \frac{1}{4 g^2_{ym} R^2} Tr \int d\Omega ( K^a_i K^b_j
K^c_i K^d_j F_{ab} F_{cd} + 2i K^a_i K^b_j F_{ab} \epsilon_{ijk}
\frac{x_k}{R} \phi \nonumber \\
& + & 2 K^a_i K^b_i ( D_a \phi ) ( D_b \phi ) - 2 \phi^2 )
\nonumber \\
& - & \frac{3}{2 g^2_{ym} R^2} Tr \int d\Omega ( i \epsilon_{ijk}
K^a_i K^b_j F_{ab} \frac{x_k}{R} \phi - \phi^2 ) \nonumber \\
& = & - \frac{1}{4 g^2_{ym} R^2} Tr \int d\Omega ( F_{ab} F^{ab} +
8i \frac{\epsilon^{ab}}{\sqrt{g}} F_{ab} \phi + 2 ( D_a \phi ) (
D^a \phi ) - 8 \phi^2 ) \nonumber \\
S_1 & = & \frac{1}{g^2_{ym} R^2} Tr \int d\Omega ( i
\epsilon_{ijk} K^a_i K^b_j F_{ab} \frac{x_k}{R} \phi - \phi^2 ) =
\nonumber \\
& = & \frac{1}{g^2_{ym} R^2} Tr \int d\Omega ( \frac{i
\epsilon^{ab}}{\sqrt{g}} F_{ab} \phi - \phi^2 ) \label{326}
\end{eqnarray}

where $ \epsilon^{ab}$ is defined as $\epsilon^{\theta \phi} = 1
$.

Since we are interested in the $U(1)$ case, we finally find

\begin{eqnarray}
S(\lambda) = S_0 + \lambda S_1 = - \frac{1}{4 g^2_{ym} R^2} \int
d\Omega [ F_{ab} F^{ab} - 2 \partial_a \phi \partial_a \phi + ( 8
- 4 \lambda ) ( \frac{i\epsilon^{ab}}{\sqrt{g}} F_{ab} \phi -
\phi^2 ) ] . \nonumber \\
& \ & \label{327} \end{eqnarray}

As in the Bogomolnyi trick, we can isolate an equation of motion
directly at the level of action as a quadratic term

\begin{eqnarray}
S( \lambda) & = & - \frac{1}{4 g^2_{ym} R^2 } \int d \Omega [ (
F_{ab} + ( 4 - 2\lambda ) i \epsilon_{ab} \phi \sqrt{g} ) (F^{ab}
+ ( 4 - 2\lambda ) i \epsilon^{ab} \frac{\phi}{\sqrt{g}} )
\nonumber \\
& - & 2 \partial_a \phi \partial_a \phi + [ 2 {( 4 - 2 \lambda
)}^2 - 4 ( 2-\lambda )] \phi^2 ] \label{328}
\end{eqnarray}

from which the resulting equations of motion are

\begin{eqnarray}
& \ & F_{ab} + ( 4 - 2 \lambda ) i \epsilon_{ab} \phi \sqrt{g} = 0
\nonumber \\
& \ & \partial^a \partial_a \phi = L_i L_i \phi = [ {( 4-2\lambda
)}^2 - 2 ( 2-\lambda ) ] \phi . \label{329} \end{eqnarray}

We can recognize in the second equation the eigenvalue problem for
the $L_i L_i$ operator which admits nontrivial solutions if and
only if

\beq L_i L_i \phi = l ( l+1 ) \phi \label{330} \eeq

the coefficient in front of $\phi$ in the second member is equal
to $l( l+1 )$. By imposing that

\beq l ( l+1 ) = {( 4 - 2 \lambda )}^2 - 2 ( 2-\lambda )
\label{331} \eeq

we find that the only classical models which admit nontrivial
solutions are for

\beq \lambda = \frac{3-l}{2} \ \ \ \ {\rm or } \ \ \ \ \lambda =
\frac{l+4}{2} . \label{332} \eeq

The case $l=0$, i.e. a constant scalar field $\phi$ can be reached
with the method outlined in the Appendix.

From the formula (\ref{223}), the scalar field $\phi(\Omega)$ is a
combination of $Y^m_2$ spherical harmonics, and we can verify that
at $l=2$ $\lambda = \half$.

Therefore, one equation of motion is surely satisfied. To check
also the other equation we need to compute $F_{\theta\phi}$

\begin{eqnarray}
F_{\theta\phi} & = & - i ( \partial_\theta b_{\phi} -
\partial_\phi b_{\theta} ) = \nonumber \\
& - & 3i sin \theta R [ 2 \lambda_0 \sqrt{ ( 1 -
\sqrt{\frac{2}{3}} )  ( 1 + 2\sqrt{\frac{2}{3}} )} sin \theta cos
\theta ( sin \phi - cos \phi ) \nonumber \\
& + & \lambda_0 \sqrt{\frac{2}{3}} sin \theta ( 2 cos^2 \theta -
sin^2 \theta  - 2 \lambda_0 ( 1 - \sqrt{\frac{2}{3}} ) sin^2
\theta cos \phi sin \phi ] = \nonumber \\
& = & - 3i sin \theta \phi( \Omega ) \epsilon_{\theta\phi}
\label{333}
\end{eqnarray}

which satisfies the other equations of motion for $\lambda =
\half$.

Therefore our solution is a noncommutative soliton for the fuzzy
sphere which, in the $N \rightarrow \infty$ limit, corresponds to
a classical soliton for the sphere with the classical action
$S(\half)$.

To find noncommutative solitons with higher spherical harmonics
$l>2$ one should explore the neighborhood of the models

\beq \lambda = \frac{3-l}{2} \label{334} \eeq

by starting from classical solutions of matrix model obtained with
more general ansatze:

\beq A_i = A_i^{j_1,j_2,..,j_{l-1}} L_{j_1} ... L_{j_{l-1}} .
\label{335} \eeq

\section{Computation of the action}

To finish our verification, let us compare the action $S(\lambda)$
evaluated at the solution $A_i$ with the action evaluated on the
classical soliton on the sphere.

Recall that the computation of the action $S(\lambda)$ on the
general class of solutions (\ref{29}), (\ref{210}), (\ref{211}) is
made of three pieces:

\begin{eqnarray}
& \ & Tr [ A^i, A^j ][ A^i, A^j ] = - \frac{8}{3} \rho^4 ( 1 -
\lambda ) ( 4 \lambda^2 - \lambda +1 ) Tr ( \hat{L}_m \hat{L}_m )
\nonumber \\
& \ & - \frac{2}{3} i \lambda \rho Tr \epsilon_{ijk} A^i A^j A^k =
\frac{8}{3} \lambda^2 ( 1-\lambda ) \rho^4 Tr ( \hat{L}_m
\hat{L}_m ) \nonumber \\
& \ & \rho^2 ( 1- \lambda ) Tr A^i A^i = \frac{4}{3} \rho^4 (
1-\lambda ) ( \lambda^2 - \lambda + 1 ) Tr ( \hat{L}_m \hat{L}_m )
. \label{41}
\end{eqnarray}

The total evaluation of the action $S(\lambda)$ is therefore given
by:

\beq S(\lambda)|_{tot} = S_0 + \lambda S_1 = - \frac{1}{3 g^2}
\rho^4 ( 1-\lambda ) ( 2 - 2 \lambda + 4 \lambda^2 ) Tr (
\hat{L}_m \hat{L}_m ) . \label{42} \eeq

This action contains a part which is divergent in the $N
\rightarrow \infty$ limit, and the remaining part which is finite.
Subtracting the divergent part we find

\beq S(\lambda)|_{finite} = S(\lambda)|_{tot} - S(\lambda)|_{div}
= \frac{4 \rho^4}{3 g^2} {(\lambda -\half )}^3 Tr ( \hat{L}_i
\hat{L}_i ) . \label{43} \eeq

By introducing the models $\lambda = \half + \rho \lambda_0$,
considered in this paper, we find:

\beq S(\lambda)|_{finite} = \frac{4}{3 g^2} \rho^7 \lambda^3_0 Tr
( \hat{L}_i \hat{L}_i )  = \frac{16 \pi \lambda^3_0}{3 R^2
g^2_{ym}} \rho \rightarrow 0 \ \ \ \ { {\rm for } \ \ \ \ N
\rightarrow \infty } \label{44} \eeq

i.e. the finite value of the action $S(\lambda)$ vanishes as
$\rho$ in the $N \rightarrow \infty$ limit.

Before comparing this null result with the value of the classical
action at the classical soliton, we shall compute the possible
finite value of the subleading term into the fluctuations, i.e.
terms of order $\rho$ with respect to the classical solitons, to
the action $S(\lambda)$.

For example, let us take the term

\beq Tr [ A^i, A^j ][A^i, A^j ] = - \frac{8}{3} \rho^4 ( 1-\lambda
) ( 4\lambda^2 - \lambda +1 ) Tr ( \hat{L}_i \hat{L}_ ) .
\label{45} \eeq

By substituting the value $ \lambda = \half + \rho \lambda_0$, the
total contribution, divergent plus finite, is given by

\beq Tr [ A^i, A^j ][A^i, A^j ]|_{total} = - \frac{8}{3} \rho^4 (
\frac{3}{4} - \rho^2 \lambda^2_0 ) Tr ( \hat{L}_i \hat{L}_i ) .
\label{46} \eeq

However the contribution of the leading fluctuations $a_i$, which
define the classical solitons, is different

\beq Tr [ A^i, A^j ][A^i, A^j ]|_{soliton} = - \frac{8}{3} \rho^4
( \frac{3}{4} + \rho^2 \lambda^2_0 ) Tr ( \hat{L}_i \hat{L}_i ) .
\label{47} \eeq

The difference is given by the subleading fluctuations
$\tilde{a}_i \sim O(\rho)$, which vanish in the classical limit,

\beq Tr [ A^i, A^j ][A^i, A^j ]|_{sub} = \frac{16}{3} \rho^6
\lambda^2_0
 Tr ( \hat{L}_i \hat{L}_i )  . \label{48} \eeq

Analogously the other two pieces of action, evaluated on the total
solution (\ref{223}), on the leading fluctuations (\ref{320}) and
on the subleading ones lead to

\begin{eqnarray}
& \ & \epsilon^{ijk} Tr A^i [ A^j, A^k ]|_{total} = 2i \rho^3 ( 1-
4 \lambda^2_0 \rho^2 ) Tr ( \hat{L}_i \hat{L}_i ) \nonumber \\
& \ & \epsilon^{ijk} Tr A^i [ A^j, A^k ]|_{soliton} = 2i \rho^3 (
1-
2 \lambda^2_0 \rho^2 ) Tr ( \hat{L}_i \hat{L}_i ) \nonumber \\
& \ & \epsilon^{ijk} Tr A^i [ A^j, A^k ]|_{sub} = - 4i \lambda^2_0
\rho^5
Tr ( \hat{L}_i \hat{L}_i ) \nonumber \\
& \ & Tr A^i A^i |_{total} = \frac{4}{3} \rho^2 [ \frac{3}{4} +
\lambda^2_0 \rho^2 ] Tr ( \hat{L}_i \hat{L}_i ) \nonumber \\
& \ & Tr A^i A^i |_{soliton} = \frac{4}{3} \rho^2 [ \frac{3}{4} +
2 \lambda^2_0 \rho^2 ] Tr ( \hat{L}_i \hat{L}_i ) \nonumber \\
& \ & Tr A^i A^i |_{sub} = - \frac{4}{3} \rho^4 \lambda^2_0 Tr (
\hat{L}_i \hat{L}_i ) . \label{49}
\end{eqnarray}

Fortunately, these finite contributions of the subleading
fluctuations cancel out from the action $S(\lambda)$. In fact for
$S_0$ and $S_1$ we find

\begin{eqnarray}
& \ & S_0 \propto ( \frac{1}{4}\frac{16}{3} \rho^6 - \frac{4}{3}
\rho^6 ) \lambda^2_0 Tr ( \hat{L}_i \hat{L}_i ) = 0 \nonumber \\
& \ & S_1 \propto ( \frac{4}{3} \rho^6 - \frac{4}{3} \rho^6 )
\lambda^2_0 Tr ( \hat{L}_i \hat{L}_i ) = 0  . \label{410}
\end{eqnarray}

We conclude that, since the total action $S(\lambda)$, evaluated
on the complete solution $A_i$, is vanishing in the $N\rightarrow
\infty$ limit, and since the subleading terms give no extra
contribution, the value of the classical action on the classical
solitons must be zero.

In fact for the classical action (\ref{328}) the equation of
motion

\beq F_{ab} + 3 i \epsilon_{ab} \sqrt{g} \phi = 0 \label{411} \eeq

cancels the quadratic part of the action, while the other equation

\beq \partial^a \partial_a \phi = 6 \phi \label{412} \eeq

cancels also the other part, since the action is homogeneously
quadratic in $\phi$:

\beq S(\half)|_{soliton} = 0  . \label{413} \eeq

However the solution is clearly non trivial, since $F_{ab}$ cannot
be made equivalent to zero with a gauge transformation.
Topological arguments as Chern - classes can eventually give rise
to a quantization of the parameter $\lambda_0$, which we have left
undetermined.

\section{Conclusion}

In this paper we have searched for non trivial solutions to the
noncommutative gauge theory over the fuzzy sphere. Between all the
possible solutions of the matrix model, those which have the right
to be taken into account are those which corresponds to the
classical solitons on a sphere in the $N \rightarrow \infty$
limit.

The fluctuations, defined as the difference between the generic
solutions of the matrix model and the background ( the fuzzy
sphere ), must vanish in the $N \rightarrow \infty$ limit as
$\frac{1}{N}$, a constraint which implies a double scaling limit
in the two coupling constants of the model, $g^2$ and $\lambda$.
Therefore to find classical solitons we have to scale $\lambda $
as $ \half + \frac{\lambda_0}{N} $.

By analyzing the classical solitons we find confirmations of the
scheme proposed, since the scalar field corresponding to our
classical solution is a combination of the spherical harmonics
$l=2$. This should be expected since the classical solution of the
matrix model we started from were found by the ansatz

\beq A^i = A^i_j L^j . \label{51} \eeq

There is in fact a correspondence between the model $\lambda =
\half $ and the spherical harmonics $l=2$. By analogy, one could
search classical solitons with higher spherical harmonics $l > 2$,
and therefore define new classes of noncommutative solitons over
the fuzzy sphere. One should firstly start from more general
ansatze:

\beq A^i = A^i_{j_1,j_2,..,j_{l-1}} L^{j_1,j_2,..,j_{l-1}}
\label{52} \eeq

and determine the corresponding classical solutions of the matrix
model. Then it should happen that these solutions are a smooth
deformation of a fuzzy sphere around the models $\lambda =
\frac{l-3}{2}$; finally to find the classical solitons with higher
spherical harmonics one should scale $\lambda$ as $\frac{l-3}{2} +
\frac{\lambda_0}{N}$.

The solitons we have found are different from the monopoles, which
are the standard solitons of gauge theory on the sphere. However
the classical theory defined from the matrix model is more
complex, being the sum of the gauge field and scalar field
actions. In fact the value of the action on these classical
solitons is zero and it is not proportional to the soliton number
as it happens with the instantons on the plane
\cite{27}-\cite{25}-\cite{26}-\cite{28}-\cite{30}-\cite{29}-\cite{24}.
However we find another signals, typical of solitons, i.e. that
the action contains an equation of motion quadratically, as it
happens with vortices.

Finally, by keeping $\lambda = \half + \frac{\lambda_0}{N}$ with
$N$ finite, our new class of solutions determines a series of
noncommutative solitons over the fuzzy sphere which converges to a
classical soliton in the $N \rightarrow \infty$ limit.

We believe that the method here outlined is a practical procedure
to define noncommutative solitons as a fuzzy sphere, and that the
matrix model approach greatly simplify the task.

It is simpler to define the noncommutative soliton as a difference
between two solutions of the matrix model, rather than trying to
solve the equations of motion of the noncommutative gauge theory
directly. It is an open question to find the quantum contribution
of this solitons to the partition function, and again this
question greatly simplifies if posed in the matrix model approach.

\appendix
\section{Appendix}

To obtain a constant scalar field as classical solution one starts
with the ansatz

\beq A^i = f(\rho) L^i = \frac{f(\rho)}{\rho} \hat{x}^i \label{A1}
\eeq

i.e. a deformation with a scale factor of the standard fuzzy
sphere solution. By imposing the equations of motion (\ref{25}),
one obtains the following quadratic form in $f(\rho)$:

\beq f^2 ( \rho ) - \lambda \rho f( \rho ) - \rho^2 ( 1-\lambda )
= 0 \label{A2} \eeq

whose solutions are

\begin{eqnarray}
& \ & f_1(\rho) = \rho \ \ \ \ \ \ {\rm standard \ fuzzy \ sphere
}
\nonumber \\
& \ & f_2(\rho) = ( \lambda-1) \rho \ \ \ \ \ {\rm rescaled \
fuzzy \ sphere } . \label{A3}
\end{eqnarray}

The case $\lambda = -1$  of the rescaled fuzzy sphere coincides
with the case $\lambda = -1$ of the class of solutions considered
in the paper.

Let us notice that for $\lambda=2$ the two solutions coincide and
the value of the classical action $S(\lambda)$ around $\lambda = 2
$ is

\begin{eqnarray}
& \ & S(\lambda)|_2 =  \frac{\rho^4}{2 g^2} {(\lambda-1)}^3 ( 1-
\frac{\lambda}{3} ) Tr ( \hat{L}_i \hat{L}_i ) \ \ \ \ \
S(\lambda)|_{{\rm fuzzy \ sphere }} = - \frac{\rho^4}{2 g^2} (
\half
- \frac{\lambda}{3}) Tr ( \hat{L}_i \hat{L}_i ) \nonumber \\
& \ & S(\lambda)|_2 - S(\lambda)|_{{\rm fuzzy \ sphere }}  =  -
\frac{\rho^4}{6 g^2} \lambda {( \lambda - 2 )}^3 Tr ( \hat{L}_i
\hat{L}_i ) . \label{A4}
\end{eqnarray}

By expanding $\lambda = 2 + \rho \lambda_0$ one then builds , in
the same way as considered in the paper, noncommutative solitons
which tend to a constant scalar field $ \phi = c$ and null field
strength $F_{ab} = 0$, in the classical limit.


\begin{thebibliography}{999}

\bibitem{1} E. Witten, " Noncommutative geometry and string field
theory ", Nucl. Phys. {\bf B268} (1986) 253.
\bibitem{2} N. Seiberg and E. Witten, " String Theory and
Noncommutative geometry ", JHEP {\bf 9909} (1999) 032,
hep-th/9908142.
\bibitem{3} A. Connes, M. R. Douglas and A. Schwarz, "
Noncommutative geometry and Matrix theory: compactification on
Tori ", JHEP {\bf 9802} (1998) 003, hep-th/9711162.
\bibitem{8} J. Ambjorn, Y.M. Makeenko, J. Nishimura and R. J.
Szabo, " Finite N Matrix models of noncommutative gauge theory ",
JHEP {\bf 9911} (1999) 029, hep-th/9911041.
\bibitem{9} J. Ambjorn, Y.M. Makeenko, J. Nishimura and R. J.
Szabo, " Non perturbative Dynamics of noncommutative gauge theory
", Phys. Lett. {\bf B480} (2000) 399.
\bibitem{10} J. Ambjorn, Y.M. Makeenko, J. Nishimura and R. J.
Szabo, " Lattice gauge fields and discrete noncommutative
Yang-Mills Theory ", JHEP {\bf 0005} (2000) 023, hep-th/0004147.
\bibitem{11} T. Eguchi and K. Kawai, " Reduction of dynamical
degrees of freedom in the large N gauge theory ", Phys. Rev. Lett.
{\bf 48} (1982) 1063.
\bibitem{12}  A. Gonzales-Arroyo and M. Okawa, " The twisted
Eguchi-Kawai model: a reduced model for large N lattice gauge
theory ", Phys. Rev. {\bf D27} (1983) 2397.
\bibitem{13} N. Ishibashi, H. Kawai, Y. Kitazawa and A. Tsuchiya,
" A large N reduced model as superstring ", Nucl. Phys. {\bf B498}
(1997) 467, hep-th/9612115.
\bibitem{4} S. Iso, Y. Kimura, K. Tanaka and K. Wakatsuki,
" Noncommutative gauge theory on fuzzy sphere from Matrix model ",
Nucl. Phys. {\bf B604} (2001) 121, hep-th/0101102.
\bibitem{5} R.C. Myers, "Dielectric branes", JHEP {\bf 9912}
(1999) 022, hep-th/9910053.
\bibitem{6} Y. Kimura, " Noncommutative gauge theories on Fuzzy
sphere and Fuzzy torus from Matrix model ", Prog. Theor. Phys.
{\bf 106} (2001) 445, hep-th/0103192.
\bibitem{7} D. Berenstein, J. M. Maldacena, H. Nastase,
" Strings in flat space and PP waves from N=4 Superyang-mills " ,
JHEP {\bf 0204} (2002) 013, hep-th/0202021.
\bibitem{32} P. Valtancoli, "Stability of the fuzzy sphere solution from matrix model,
hep-th/0206075.
\bibitem{14} J. Madore, " The Fuzzy sphere ", Class. Quantum Grav.
{\bf 9} (1992) 69.
\bibitem{15} U. Carow-Watamura and S. Watamura, " Noncommutative
geometry and gauge theory on Fuzzy sphere ", Comm. Math. Phys.
{\bf 212} (2000) 395, hep-th/9801195.
\bibitem{16} U. Carow-Watamura and S. Watamura, " Differential
calculus on fuzzy sphere and scalar field ", Int. J. Mod. Phys.
{\bf A13} (1998) 3235, q-alg/9710034.
\bibitem{17} U. Carow-Watamura and S. Watamura, " Chirality and
Dirac operator on Noncommutative sphere ". Comm. Math. Phys. {\bf
183} (1997) 365, hep-th/9605003.
\bibitem{18} C. Klimcik, " Gauge theories on the noncommutative
sphere ", Comm. Math. Phys. {\bf 199} (1998) 257, hep-th/9710153.
\bibitem{19} H. Grosse and P. Presnajder, " The Dirac operators on
the fuzzy sphere ", Lett. math. Phys. {\bf 33} (1995) 171.
\bibitem{20} S. Baez, A.P. Balachandran, S. Vaidya, " Monopoles
and Solitons in fuzzy physics ", Comm. Math. Phys. {\bf 208}
(2000) 787,hep-th/9811169.
\bibitem{21} A.P. Balachandran and S. Vaidya, " Instantons
and chiral anomaly in fuzzy physics ", hep-th/9910129.
\bibitem{22} A. P. Balachandran, X.Martin and D. O' Connor, "
Fuzzy actions and their continuum limits ", hep-th/0007030.
\bibitem{23} P. Valtancoli, " Projectors for the fuzzy sphere "
Mod.Phys.Lett. {\bf A16} (2001) 639, hep-th/0101189.
\bibitem{31} S. Vaidya, " Scalar multi-solitons on the fuzzy
sphere ", JHEP {\bf 0201} (2002) 011, hep-th/0109102.
\bibitem{27} N. Nekrasov, A. Schwarz, " Instantons on
noncommutative $R^4$ and $(2,0)$ superconformal six dimensional
theory ", Comm. Math. Phys. {\bf 128} (1998) 689.
\bibitem{25} D. J. Gross, N. A. Nekrasov, " Solitons in
noncommutative gauge theory ", JHEP {\bf 0103} (2001)
044,hep-th/0010090.
\bibitem{26} D. J. Gross, N. A. Nekrasov, " Monopoles and strings
in noncommutative gauge theory ", JHEP {\bf 0007} (2000) 034,
hep-th/0005204.
\bibitem{28} K. Furuuchi, " Instantons on noncommutative $R^4$ and
projection operators ", Prog. Theor. Phys. {\bf 103} (2000) 1043,
hep-th/9912047.
\bibitem{30} L. D. Paniak, R. J. Szabo, " Instanton expansion of
noncommutative gauge theory in two dimensions ", hep-th/0203166.
\bibitem{29} J. A. Harvey, P. Kraus, F. Larsen, " Exact
noncommutative solitons ", JHEP {\bf 0012} (2000) 024,
hep-th/0010060.
\bibitem{24} P. Krauss and M. Shigemori, " Noncommutative
Instantons and the Seiberg-Witten Map ", JHEP {\bf
 0206} (2002) 034, hep-th/0110035.
\end{thebibliography}
\end{document}